\documentclass[12pt,a4paper]{article}
\usepackage[utf8]{inputenc}
\usepackage{color}
\usepackage{parskip}
\usepackage{enumitem}
\usepackage{amsmath,amscd}
\usepackage{stmaryrd}
\usepackage{mathrsfs}
\usepackage{amsfonts}
\usepackage{amssymb,scalerel,stackengine}
\usepackage[most]{tcolorbox}
\newtcbox{\othermathbox}[1][]{nobeforeafter, math upper, tcbox raise base, 
          enhanced, rounded corners, colback=black!5, colframe=black}
\usepackage{empheq}
\usepackage[margin=1in]{geometry}
\usepackage{graphicx}
\usepackage[leftcaption]{sidecap}
\usepackage{caption}
\usepackage{subcaption}
\usepackage{here}
\usepackage[colorlinks=true,allcolors=blue]{hyperref}
\usepackage{cite}
\let\OLDthebibliography\thebibliography
\renewcommand\thebibliography[1]{
  \OLDthebibliography{#1}
  \setlength{\parskip}{0pt}
  \setlength{\itemsep}{4.85pt plus 0.3ex}
}

\newcommand\beq{\begin{equation}}
\newcommand\ee{\end{equation}}
\newcommand\cO{{\cal O}}

\newcommand\der{\partial}
\newcommand\ie{\textit{i.e.}\ }
\newcommand\eg{\textit{e.g.}\ }

\def\hi{{\hat i}}
\def\hj{{\hat j}}

\def\hm{{\hat m}}
\def\hn{{\hat n}}
\def\ha{{\hat a}}
\def\hb{{\hat b}}

\def\hr{{\hat r}}
\def\mn{{\mu\nu}}

\def\de{\delta}

\def\pd{\partial}

\def\cd{\nabla}
\def\eps{\epsilon}
\def\veps{\varepsilon}
\def\cL{\mathcal{L}}
\def\cJ{\mathcal{J}}
\def\cM{\mathcal{M}}

\def\cC{\mathcal{C}}

\def\cF{\mathcal{F}}

\def\cP{\mathcal{P}}

\def\bW{\boldsymbol{\Omega}}

\def\cO{\mathcal{O}}

\def\bl{\boldsymbol{\ell}}
\def\bk{\boldsymbol{k}}

\def\bS{\boldsymbol{S}}

\def\bn{\boldsymbol{n}}
\def\bq{\boldsymbol{q}}
\def\bth{\boldsymbol{\theta}}

\def\th{\theta}

\def\tl{\widetilde}

\newcommand\be[1]{{\boldsymbol{e}_{\hat{#1}}}}
\newcommand\bff[1]{{\boldsymbol{f}_{\hat{#1}}}}
\newcommand\e[2]{{e_{\hat{#1}}^{\;\;#2}}}
\newcommand\E[2]{{\zeta_{\hat{#1}}^{\;\;#2}}}
\newcommand\barE[2]{{\overline{\zeta}_{\hat{#1}}^{\;\;#2}}}
\def\bu{{\boldsymbol u}}

\title{\bf Gyroscopic Gravitational Memory}
\author{Ali Seraj$^{\boldsymbol+}$ and Blagoje Oblak$^{\boldsymbol\times}$\\[1em]
{\small $^{\boldsymbol+}$ Centre for Gravitational Waves, Université Libre de Bruxelles,}\\[-.2em]
{\small International Solvay Institutes, CP 231, B-1050 Brussels, Belgium.}\\[.4em]
{\small $^{\boldsymbol\times}$ CPHT, CNRS, Ecole Polytechnique, IP Paris, F-91128 Palaiseau, France;}\\[-.2em]
{\small LPTHE, CNRS and Sorbonne Université, 75252 Paris Cedex 05, France.}}
\date{}

\begin{document}

\maketitle

\begin{center}
\begin{minipage}{.87\textwidth}
{\bf Abstract.} We study the motion of a gyroscope located far away from an isolated gravitational source in an asymptotically flat spacetime. As seen from a local frame tied to distant stars, the gyroscope precesses when gravitational waves cross its path, resulting in a net `orientation memory' that carries information on the wave profile. At leading order in the inverse distance to the source, the memory consists of two terms: the first is linear in the metric perturbation and coincides with the spin memory effect, while the second is quadratic and measures the net helicity of the wave burst. Both are closely related to symmetries of the gravitational radiative phase space at null infinity: spin memory probes superrotation charges, while helicity is the canonical generator of local electric-magnetic duality on the celestial sphere.
\end{minipage}
\end{center}

\newpage
\tableofcontents

\newpage
\section{Introduction}

The direct observation of gravitational waves from a merger of binary black holes \cite{LIGOScientific:2016aoc} marked the birth of an immensely exciting scientific era: gravitational radiation now provides a new window to observe the universe, complementing the electromagnetic signals traditionally used in astrophysics. However, multi-messenger astronomy requires a precise understanding and modelling of gravitational waves, which is made difficult by the nonlinear nature of general relativity. It is therefore of interest to find new, potentially measurable quantities sensitive to gravitational wave profiles. The goal of this paper is to describe one such observable affecting what is, perhaps, the simplest measuring device of astronomers: a spinning gyroscope \cite{Herrera:2000uh,Bini:2000xj,ValienteKroon:2001pc}.

A prominent manifestation of gravitational nonlinearity appears in so-called hereditary effects \cite{Blanchet:2013haa}, which depend on a system's entire past history. In particular, \textit{gravitational wave memory} \cite{Zeldovich:1974gvh,Braginsky:1985vlg,braginsky1987gravitational,Christodoulou:1991cr,Blanchet:1992br} is a permanent net change of certain metric components whose simplest consequence is displacement memory, \ie the permanent change of distance between two nearby freely-falling test masses after the passage of gravitational waves. The detection of such phenomena is hampered by the poor sensitivity of gravitational wave detectors at low frequencies, but an encouraging fact is that memory is a \textit{Newtonian} addition to the oscillatory waveforms emitted by bounded gravitational sources \cite{Wiseman:1991ss}.\footnote{More precisely, while memory originates from the nonlinear gravitational interaction, its accumulation over the coalescence period leads to a Newtonian correction to the waveform, \ie one that does not involve additional powers of Newton's constant with respect to the leading oscillatory waveform.} As a result, the effect may realistically be large for binary systems \cite{Mitman:2020bjf,Mitman:2020pbt,Mitman:2021xkq}, and could be measured in future experiments: see \eg \cite{Favata:2010zu,Lasky:2016knh,Boersma:2020gxx}. Here we shall argue that gyroscopes similarly display a potentially observable `orientation memory' sensitive to gravitomagnetic components of the radiating field.

Indeed, there exists by now a plethora of distinct manifestations of gravitational memory, ranging from the aforementioned displacement memory \cite{Zeldovich:1974gvh,Braginsky:1985vlg,braginsky1987gravitational} to the kick (velocity) memory effect caused by planar waves \cite{Zhang:2017rno,Zhang:2017geq} or compact sources \cite{Flanagan:2018yzh,Divakarla:2021xrd}, spin memory in Sagnac interferometers \cite{Pasterski:2015tva,Mao:2018xcw}, and novel memory effects in modified theories of gravity \cite{Tahura:2020vsa,Hou:2020tnd,Seraj:2021qja,Tahura:2021hbk}. The unifying thread \cite{Strominger:2014pwa,Strominger:2015bla} underlying these observables is the existence of asymptotic Bondi-Metzner-Sachs (BMS) symmetries of gravity \cite{Bondi:1962px,Sachs2} and their various extensions \cite{Barnich:2009se,Barnich:2010eb,Campiglia:2014yka,Freidel:2021fxf}---an active field of research on its own.

We will see that gyroscopic memory is related to gravitational symmetries as well. Indeed, the precession rate of a gyroscope in a gravitational wave background is a sum of two terms, respectively linear and quadratic in the metric perturbation, at leading order in the inverse distance to the source. The coefficients of this sum turn out to be just such that the precession rate coincides with the `covariant dual mass aspect' \cite{Freidel:2021qpz} whose linear part is the boundary Noether current of dual supertranslations in an extended version of the standard BMS group \cite{Godazgar:2018qpq,Godazgar:2018dvh,Kol:2019nkc,Godazgar:2019dkh,Godazgar:2020gqd,Godazgar:2020kqd,Oliveri:2020xls,Kol:2020ucd}; it is also the canonical generator of nontrivial asymptotic frame rotations in the first-order formulation of general relativity \cite{Godazgar:2022pbx}. Since the gyroscopic memory effect is obtained by integrating the precession rate over time, it similarly contains two contributions: the linear one coincides with the spin memory effect \cite{Pasterski:2015tva,Nichols:2017rqr} and is related to superrotation charges through flux-balance equations for angular momentum, while the nonlinear flux term measures the helicity of the passing waves and generates canonical duality transformations in radiative phase space. Remarkably, this linear $+$ quadratic form of the precession rate seems to be universal in gauge theory: it also occurs in the case of magnetic dipoles subjected to electromagnetic waves \cite{Oblak:2023axy}, although we will not discuss this setup here.

To conclude this introduction, a comment is called for regarding the magnitude of precession as a function of the observer's distance to the source. While usual displacement memory is proportional to the inverse distance, gyroscopic memory (like spin memory) is proportional to its \textit{square}: it is a weaker effect. This is not surprising, as even the best known precession effect in gravitational backgrounds with angular momentum---namely Lense-Thirring precession (see e.g. \cite[\S40.7]{Misner:1973prb})---involves the \textit{cube} of the inverse distance. In that respect, our computation generalizes the Lense-Thirring effect to radiative spacetimes and `corrects' it by a dominant, overleading term at large distances.

The paper is organized as follows. Section \ref{se2} is devoted to a lightning review of the Bondi formalism for asymptotically flat gravitational fields, and introduces asymptotic geodesics therein. These are then used in section \ref{se3} to describe the kinematics of freely-falling (or mildly accelerated) gyroscopes with respect to a natural choice of local tetrad---namely one that is tied to distant stars and obtained by suitably rotating a `source-oriented' frame where one vector points towards the origin of radiation. Finally, section \ref{se4} displays the precession rate and ensuing gyroscopic memory for freely-falling observers, whereupon the result is related to dual asymptotic symmetries and flux-balance equations of gravitational data. Static observers are also briefly considered, and their gyroscopic memory effects are compared to earlier statements in the literature, such as asymptotic vorticity \cite{Herrera:2000uh,Herrera:2013poa} and spin memory \cite{Pasterski:2015tva}. We conclude in section \ref{se5} with a summary and a discussion of potential follow-ups. Note that an abridged version of this material was recently published in \cite{Seraj:2022qyt}.

\section{Radiative asymptotically flat metrics}
\label{se2}

This preliminary section serves to set up our notation, review the Bondi formalism for asymptotically Minkowskian metrics, and derive geodesics near null infinity. These geodesics will eventually be the worldlines of freely-falling gyroscopes in section \ref{se3}.

\subsection{Off-shell metric in Bondi gauge}
\label{seconv}

\begin{SCfigure}[2][t]
\centering
\includegraphics[width=.3\textwidth]{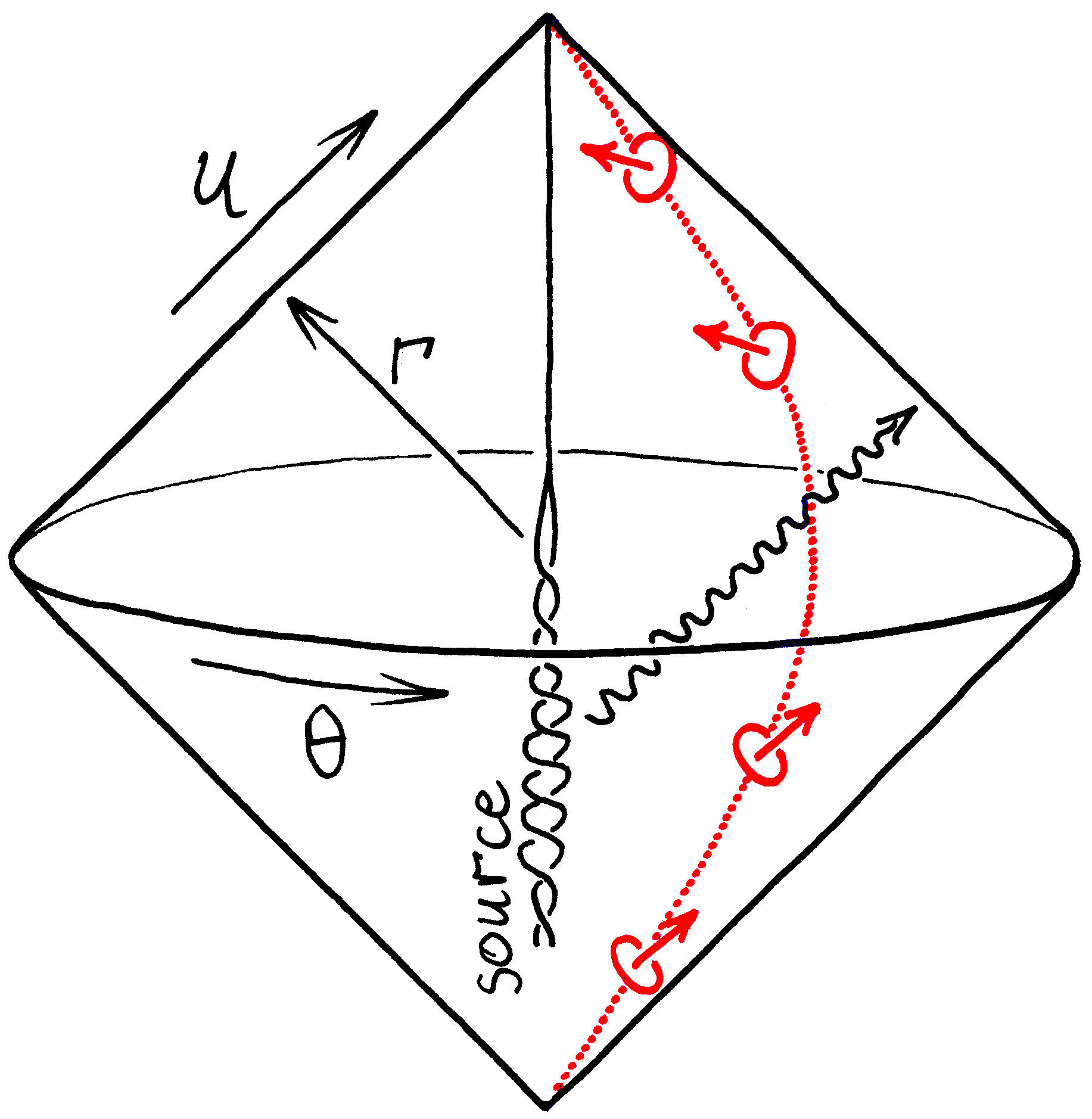}
\caption{A Penrose diagram of some asymptotically flat spacetime, including Bondi coordinates $(u,r,\th^a)$ in terms of which the (off-shell) metric takes the form \eqref{e2}. Throughout this work, we assume that some source of gravitational radiation is located near the origin ($r=0$). The resulting gravitational waves cross the worldline of a gyroscope (red) and cause its orientation to change with respect to a tetrad built in section \ref{se3}. This change of orientation may be seen as a gravitomagnetic memory effect and is studied in detail in section \ref{se4}.}
\label{fiZZ}
\end{SCfigure}

Consider a Lorentzian spacetime manifold endowed with (retarded) Bondi coordinates $(u,r,\th^a)$, where $u$ is retarded time, $r$ is a radius, and $\th^a$ ($a=1,2$) are coordinates on a celestial sphere at future null infinity (see fig.\ \ref{fiZZ}). The manifold carries a metric $ds^2=g_{\mu\nu}dx^{\mu}dx^{\nu}$ whose components are taken to satisfy the Bondi gauge conditions
\beq
\label{bog}
g_{rr}=g_{ra}=0,
\qquad
\pd_r\det\left(r^{-2}g_{ab}\right)=0.
\ee
Any such metric is commonly written as 
\beq
\label{e2}
ds^2
=
-e^{2\beta}\big(F d u^{2}+2 du\,dr\big)
+r^2\gamma_{ab}\Big(d\th^a-\frac{\,U^{a}}{r^2} du\Big)\Big(d\th^b-\frac{\,U^{b}}{r^2}du\Big)
\ee
in terms of a metric $\gamma_{ab}$ that will eventually include the effects of gravitational radiation, a vector field $U^a$ that will eventually carry information on angular momentum, and two functions $\beta,F$, the second of which will eventually sense the mass of the source. Off-shell, all these quantities are \textit{a priori} arbitrary functions on spacetime. To ensure that the metric \eqref{e2} is in fact asymptotically \textit{flat}, one imposes the extra boundary condition\footnote{Here and in the remainder of the paper, we use the single bold letter $\bth$ to denote a point on $S^2$ with coordinates $(\th^1,\th^2)$. In particular, $\bth$ is \textit{not} the polar angle of a spherical coordinate system.}
\beq
\label{BC}
\lim_{r\to \infty}\gamma_{ab}(u,r,\bth)
\equiv
q_{ab}(\bth),
\ee
where the right-hand side is the static, radius-independent metric of a unit sphere (with Ricci scalar $R[q]=2$) written in coordinates $\th^a$. Eq.\ \eqref{BC} and the last condition in \eqref{bog} then imply that a sphere at constant $(u,r)$ has area $4\pi r^2$, so $r$ measures areal distance. No further boundary conditions are needed: in vacuum Einstein gravity, eq.\ \eqref{BC} turns out to fix the on-shell radial dependence of all other variables, as we shall recall below.

The gauge conditions \eqref{bog} imply that constant $u$ hypersurfaces are light-like, as $g^{uu}=g^{\mn}\pd_\mu u\,\pd_\nu u=0$. Their normal vector $\bl\equiv -g^{\mn}\pd_\nu u\,\pd_\nu= e^{-2\beta}\pd_r$ satisfies $\bl\cdot \nabla \bl=0$, thus representing a congruence of outgoing affine null geodesics. Indeed, we are interested in asymptotically flat spacetimes with a gravitational source near the origin\footnote{We assume that the matter stress tensor has compact support in the bulk. Delocalized (\eg electromagnetic) sources would require, instead, additional fall-offs for the stress tensor (see \eg \cite{Flanagan:2015pxa}).}---the initial center of mass of the source is assumed to be at $r=0$---so $\bl$ is tangent to null rays emitted from the source, and integral curves of $\bl$ eventually reach future null infinity (the region $r\to\infty$ where all other coordinates are kept finite). One should thus think of the large-$r$ region as the typical location of detectors of gravitational waves. Indeed, it is the tangent vector $\bl$ that will eventually be used in section \ref{se3} to build a source-oriented tetrad carried by observers far away from the source.

\newpage
\subsection{On-shell metrics}

The Bondi framework naturally suggests solving Einstein's equations perturbatively in $1/r$, \ie as an expansion near null infinity. Vacuum dynamics thus constrains the arbitrary functions of \eqref{e2} in the form of asymptotic expansions \cite{Flanagan:2015pxa}
\begin{align}
\label{fex}
F
&=
1-\frac{2 m}{r}-\frac{2F_2}{r^2}+\cO(r^{-3}),\\
\label{bex}
\beta
&=
\frac{\beta_{2}}{r^2}+\cO(r^{-3}),\\
\label{gex}
\gamma_{ab}
&=
q_{ab}+\frac{1}{r} C_{ab}+\frac{1}{r^{2}} D_{ab}+\cO(r^{-3}),\\
\label{uex}
{U}^{a}
&=
\mathring{U}^{a}+\frac{1}{r}\left[-\frac{2}{3} L^{a}+\frac{1}{16} D^{a}(C_{bc} C^{bc})
+\frac{1}{2} C^{ab} D^{c} C_{bc}\right]+\cO(r^{-2}),
\end{align}
where indices $a,b$ on celestial spheres are raised and lowered with the static metric \eqref{BC}, $D$ denotes the spherical Levi-Civita connection, and each coefficient of $1/r^n$ is some function of $(u,\bth)$, as follows. First, the subleading metric correction $C_{ab}$ in \eqref{gex} is the Bondi shear \cite{Christodoulou:1991cr}, measuring (as its name indicates) the shear of outgoing null geodesic congruences near infinity. It is a symmetric and traceless tensor that depends \textit{freely} on $(u,\bth)$; in fact, it is the only genuinely free data of on-shell metrics, and it will be the key object contributing to precession in sections \ref{se3}--\ref{se4}. Along with initial conditions on mass and angular momentum (see below), it determines all other subleading coefficients in \eqref{fex}--\eqref{uex} according to
\begin{align}
\label{e9}
F_2
&=
-\frac{1}{32}C^2
-\frac{1}{6}\left(D_{a} L^{a}\right)-\frac{1}{8}\left(D_{a} C^{a b}\right)\left(D_{d} C_{b}^{d}\right),\\
\label{e10}
\beta_2
&=
-\frac{1}{32}C^2,
\qquad 
D_{ab}
=
\frac14 q_{ab}C^2,
\qquad 
\mathring{U}^{a}
=
-\frac12 D_bC^{ab},
\end{align}
where $C^2\equiv C_{ab} C^{ab}$ for brevity. The only quantities that are \textit{not} directly fixed by shear are the Bondi mass aspect $m(u,\bth)$ of eq.\ \eqref{fex}, and the angular momentum aspect $L^a(u,\bth)$ in \eqref{uex}--\eqref{e9} (respectively measuring densities of energy and angular momentum on celestial spheres): their initial configuration at some time $u_0$ is arbitrary, but their time evolution is otherwise given by the news tensor
\beq
N_{ab}\equiv\pd_uC_{ab}
\label{newsdef}
\ee
according to the balance equations \cite{Flanagan:2015pxa}
\beq
\label{balance}
\dot{m}
=
\frac{1}{4}D_aD_bN^{ab}-\cP,
\qquad
\dot{L}_{a}
=
D_am+\frac{1}{2}D^bD_{[a}D^cC_{b]c}
-\cJ_a,
\ee
where the dot denotes partial derivatives with respect to retarded time $u$, and we have introduced local quadratic fluxes of energy and angular momentum:
\beq
\label{fluxes}
\cP
\equiv
\frac{1}{8} N_{ab}N^{ab},
\qquad
\cJ_a
\equiv
-\frac{1}{4} D_b(N^{bc} C_{ac})-\frac{1}{2} D_bN^{bc}C_{ac}.
\ee
The balance equations \eqref{balance} will play an important role in section \ref{se4}, allowing us to relate gyroscopic memory to angular momentum fluxes. We stress once more that the time-dependence of news is arbitrary, unless a specific gravitational source has been chosen.\footnote{Suitably `generic' asymptotically flat initial data actually satisfy $N_{ab}\sim\cO(|u|^{-3/2})$ at late and early times $u\to\pm\infty$ \cite{christodoulou1993global}, but this will play no role for our purposes.}

To summarize, the on-shell form of the Bondi metric \eqref{e2} satisfies the asymptotics
\beq
\label{e13}
\begin{split}
ds^2
\sim&
-\Big(1{-}\tfrac{2m}{r}{-}\tfrac{2F_2}{r^2}\Big)du^2-2\left(1{-}\tfrac{C^2}{16r^2}\right)du\,dr
+\Big(r^2q_{ab}{+}rC_{ab}{+}\tfrac{1}{4}q_{ab}C^2\Big)d\th^a\,d\th^b\\
&+2\Big(\tfrac{1}{2}D^bC_{ab}-\tfrac{1}{r}\left[-\tfrac{2}{3}L_a+\tfrac{1}{16}\pd_a(C^2)\right]
\Big)du\,d\th^a
\end{split}
\ee
at large $r$, where the omitted components $rr$ and $ra$ vanish identically owing to \eqref{bog}. The leading actor in this expression is the asymptotic shear $C_{ab}(u,\bth)$, which fixes $F_2$ according to eq.\ \eqref{e9} and determines the news $N_{ab}$ that crucially affects the balance equations \eqref{balance} for energy and angular momentum. Subleading corrections to the metric components in \eqref{e13} turn out to be irrelevant in what follows, so we systematically neglect them.

\subsection{Asymptotic geodesics}

We conclude this section by writing down the large-$r$ asymptotic solution of the geodesic flow equation $\bu\cdot\cd\bu=0$. (This is a key preliminary for freely-falling tetrads and their spin connections built in section \ref{se3}.) For definiteness, integration constants are fixed by demanding that freely-falling observers have a vanishing velocity relative to the source at some initial time $u=u_0$. Thus, upon expressing proper velocity as
\beq
\label{e14}
\bu
=
\gamma(\pd_u+v^r\pd_r+v^a\pd_a),    
\ee
a mildly tedious but straightforward computation yields
\begin{align}
\label{geu}
\gamma
&=
1+\frac{m_0}{r}+\frac{\gamma_2}{r^2},
\qquad
\gamma_2
=
\int_{u_0}^u du' m+ \frac{1}{16}\Delta(C^2),\\
\label{ger}
v^r
&=
\frac{\Delta m}{r}-\frac{1}{r^2}\Big[\gamma_2+\Delta\Big(\frac{1}{6}D_aL^a+\frac{1}{8}(D\cdot C)^2\Big)\Big],\\
\label{gea}
v^a
&=
-\frac{1}{2r^2}D_b\Delta C^{ab}+\frac{1}{r^3}\Big(D^a\gamma_2-\frac23\Delta L^a+\frac12C^{ab}D^c \Delta C_{bc}\Big),
\end{align}
where we write $\Delta X\equiv X(u)-X(u_0)$ for any time-dependent quantity $X$, $m_0=m(u_0,\bth)$ is the initial mass aspect, and $(D\cdot C)^2\equiv D_bC^{ab}D^cC_{ac}$. Note the following elementary consistency check: in nonradiative spacetimes (where  $N_{ab}=0$, hence $\dot m=0$), the leading radial velocity \eqref{ger} contains a term $-m_0\Delta u/r^2$ caused by the usual Newtonian gravitational acceleration $\dot{v}^r=-m_0/r^2$.

As it happens, the metric \eqref{e13} along with eqs.\ \eqref{e14}--\eqref{gea} already contain enough information to reproduce the standard displacement memory effect \cite{Strominger:2014pwa}. The key lesson of eqs.\ \eqref{geu}--\eqref{gea} is indeed that freely-falling observers nearly at rest in Bondi coordinates have a velocity $\bu\sim\partial_u$. Then the $ab$ components of the metric \eqref{e13} show that the angular distance between two nearby geodesics is $(q_{ab}+\tfrac{1}{r}C_{ab})d\th^ad\th^b$, which immediately implies that the presence of radiation ($\dot C_{ab}\neq0$) generally entails a net change $\tfrac{1}{r}\Delta C_{ab}d\th^ad\th^b$ in angular distance, where $\Delta C_{ab}\equiv\int du\,\dot C_{ab}=C_{ab}(u=+\infty)-C_{ab}(u=-\infty)$. At a deeper level, this reflects a deviation of nearby geodesics near null infinity (as can be confirmed from a computation of the relevant components of the Riemann tensor). It is likely to be the most easily observable kind of gravitational memory \cite{Favata:2010zu}, in part thanks to its $\cO(1/r)$ behaviour, but we will not describe it in detail here. Instead, we shall focus on a subleading notion of memory (that will scale as $\cO(1/r^2)$) whose advantage is to be local on a celestial sphere: it does not require a comparison of nearby geodesics. This first requires a discussion of tetrads carried by freely-falling observers.

\section{Gyroscopes in local frames}
\label{se3}

This section is devoted to the kinematics of a gyroscope near null infinity in an asymptotically flat gravitational field whose metric takes the Bondi form \eqref{e2}. Accordingly, we start by reviewing the equations of motion of a spin vector relative to any local reference frame in terms of the associated spin connection. We then construct a frame tied to distant stars by first building a source-oriented tetrad, then performing suitable angle-dependent rotations. We end by displaying the spin connections associated with freely-falling observers and static observers in Bondi coordinates. Note that most of the discussion for now will be off-shell: the metric \eqref{e2} need not satisfy Einstein's equations. This is all a key prerequisite for section \ref{se4}, where the spin connection gives rise to gyroscopic precession, hence to memory effects once gravitational dynamics is taken into account.

\subsection{Gyroscopic kinematics}
\label{segyk}

Consider a (small\footnote{This drastically simplifies the problem: the spin of large freely-falling bodies generally satisfies intricate relativistic evolution equations \cite{Dixon:1970zza}, but these boil down to parallel transport for small systems.}) gyroscope with proper velocity $\bu=u^\mu\pd_\mu$ and spin $\bS=S^{\mu}\pd_\mu$ such that $\bS\cdot\bu=0$. When the gyroscope falls freely, its spin obeys the parallel transport equation $\bu\cdot\nabla\bS=0$. More generally, if external forces cause an acceleration $a^\mu=u^\nu\cd_\nu u^\mu$, the spin vector obeys Fermi-Walker transport $(\bu\cdot\nabla\bS)^\mu=(u^\mu a^\nu-u^\nu a^\mu)S_\nu$ so as to remain orthogonal to velocity \cite[chap.\ 6]{Misner:1973prb}. Now let an observer carrying the gyroscope measure its orientation with respect to some local tetrad of vectors $\be{\mu}=\e{\mu}{\nu}\pd_\nu$ that are orthonormal in the sense that $\be{\mu}\cdot \be{\nu}=\eta_{\hat\mu\hat\nu}$. The spin vector can then be written with hatted components as $\bS=S^{\hat\mu}\be{\mu}$, and the Fermi-Walker equation becomes
\beq
\label{e17}
\frac{dS^{\hat \mu}}{d\tau}
=
\big(-u^{\alpha}
\omega_\alpha^{\;\;\hat \mu \hat \nu}+u^{\hat \mu}a^{\hat \nu}\big)S_{\hat \nu} 
\ee
where $u^{\hat \mu}=e^{\hat \mu}{}_\nu u^\nu$ are the components of proper velocity in the observer's frame, while
\beq
\label{e18}
\omega_{\mu}{}^{\hat\mu \hat\nu}
\equiv
e^{\hat\mu}{}_{\alpha} \nabla_{\mu} e^{\hat \nu\alpha }
=
e^{\hat\mu}{}_{\alpha}\big(\pd_{\mu} e^{\hat\nu\alpha }+\Gamma_{\sigma \mu}^{\alpha} e^{\hat\nu\sigma }\big)
\ee
is the spin connection one-form associated with the tetrad $\{\be{\mu}\}$. (Hatted indices are raised and lowered with $\eta_{\hat{\mu}\hat{\nu}}$, plain indices with $g_{\mn}$.) Eq.\ \eqref{e17} states that the precession of gyroscopes is determined by the spin connection, so the latter is a key object. The fact that the precession rate is independent of the properties of the gyroscope (\eg moments of inertia) may be seen as a manifestation of the equivalence principle. This would \textit{not} be the case \eg for magnetic dipoles in electrodynamics \cite{Oblak:2023axy}.

In what follows, we always restrict attention to frames that are adapted to the observer in the sense that $\be{0}=\bu$ coincides with proper velocity. Then the spin vector $\bS=S^\hi\be{i}$ is purely spatial ($i=1,2,3$) and the precession equation \eqref{e17} becomes
\beq
\label{e19}
\frac{dS^{\hat{i}}}{d\tau}
=
\Omega^{\hat{i}}_{\;\hat{j}}  S^{\hat{j}},
\qquad
\Omega^{\hat{i}\hat{j}}
\equiv
-u^{\alpha}\omega_\alpha^{\;\;\hat i \hat j},
\ee
where the observer's acceleration no longer contributes. The antisymmetric tensor $\Omega_{\hat{i}\hat{j}}$ thus yields the gyroscope's precession rate. It can be dualized into a vector $\bW=\Omega^{\hat{i}}\be{i}\equiv-\tfrac{1}{2}\eps^{{\hat{i}}{\hat{j}}{\hat{k}}}\Omega_{\hat{j}\hat{k}}\be{i}$ thanks to the usual Levi-Civita symbol, so that eq.\ \eqref{e19} reads $\dot\bS=\bW\times\bS$ in terms of the Euclidean cross product.

Determining the gyroscope's motion thus requires that one choose a velocity $\bu$ and a local frame to compute the spin connection \eqref{e18}. Accordingly, we now build two tetrads: the first will be source-oriented (section \ref{sesof}), while the second will be tied to distant stars (section \ref{sestar}). As for the observer's velocity, we shall consider two cases: free fall, and static locations at constant $(r,\bth)$ in Bondi coordinates. Both types of worldlines are understood to live at large $r$, so asymptotic expansions near infinity will often be used.

\subsection{Source-oriented frame}
\label{sesof}

Regardless of whether an observer traces a geodesic or not, let their proper velocity read
\beq
\label{e21}
\bu
=
\dfrac{dx^\mu}{d\tau}\pd_{\mu}
=
\gamma\big(\pd_u+v^r\pd_r+v^a\pd_a\big)
\ee
in Bondi coordinates, where the `time dilation factor' 
\beq
\gamma
=
\frac{du}{d\tau}
=
\left[%
e^{2\beta}(F+2v^r)-r^2\gamma_{ab}(v^a-U^a/r^2)(v^b-U^b/r^2)
\right]^{-1/2}
\ee
normalizes $\bu$ with respect to the off-shell metric \eqref{e2}. The earlier geodesic velocity \eqref{e14} is a special case of \eqref{e21}, while the velocity of a static observer just reads $\bu=\gamma\pd_u$. We now build a local tetrad $\{\be{0}{},\be{r}{},\be{a}{}\}$ in a way that is compatible with the observer so that $\be{0}{}=\bu$, and adapted to the source in that $\be{r}$ is aligned with light rays emitted at the origin. To obtain $\be{r}$, recall from section \ref{se2} that $\bl=e^{-2\beta}\pd_r$ is tangent (at all orders in $1/r$) to outgoing null rays; the radial tetrad vector thus coincides with $\bl$ up to the fact that its component along $\be{0}$ is projected out, \ie
\beq
\label{e23}
\be{r}
=
\frac{1}{\gamma}\bl-\bu.
\ee
The tetrad is completed by two more vectors $\be{a}$ whose orthogonality to $\be{0},\be{r}$ yields
\beq
\label{e24}
\be{a}
=
\frac{\E{a}{a}}{r}\left[\pd_a+\gamma_{ab}\Big(r^2 v^{b}-U^{b}\Big)\bl\right]
\ee
for some linearly independent vectors $\E{a}{\!}$ tangent to the sphere. The remaining orthonormality conditions $\be{a}\cdot\be{b}=\de_{\hat a\hat b}$ then imply that these $\E{a}{\!}$'s form an orthonormal dyad with respect to the time-dependent, radius-dependent metric $\gamma_{ab}(u,r,\bth)$:
\beq
\label{e25}
\gamma_{ab}\,\E{a}{a}\,\E{b}{b}=\de_{\hat a\hat b}.
\ee
In contrast to timelike and radial frame vectors, this does not fix the angular vectors uniquely since one is free to perform local rotations of $\ha\hb$ indices that depend on all Bondi coordinates while preserving the relation \eqref{e25}. We therefore fix this ambiguity by requiring that angular tetrad vectors point in the same direction at different radii, which is to say that $(\bl\cdot\cd \be{a})^a=0$. Together with the on-shell expressions \eqref{gex}--\eqref{e10} and the identity $C_{ac}C^{bc}=\frac12 C^2\de_a^b$, this condition implies that
\begin{align}
\label{dyad-bulk}
    \E{a}{a}\sim\barE{a}{b}\bigg(\de_b^a-\frac{1}{2r}C_b{}^a+\frac{1}{16r^2}C^2\de_b^a\bigg)\,,
\end{align} 
where $\barE{a}{}(\bth)$ is a time-independent dyad on the celestial sphere such that $q_{ab}\barE{a}{a}\barE{b}{b}=\de_{\hat a\hat b}$. Fixing the dyad $\bar\zeta_{\ha}$ then determines the angular frame vectors \eqref{e24} uniquely. Of course, the dyad $\bar\zeta_{\ha}$ itself is only fixed up to local U(1) rotations on the sphere \cite{Godazgar:2022foc}; we briefly return to those at the end of section \ref{segymem}.

\begin{SCfigure}[2][t]
\centering
\includegraphics[width=.5\textwidth]{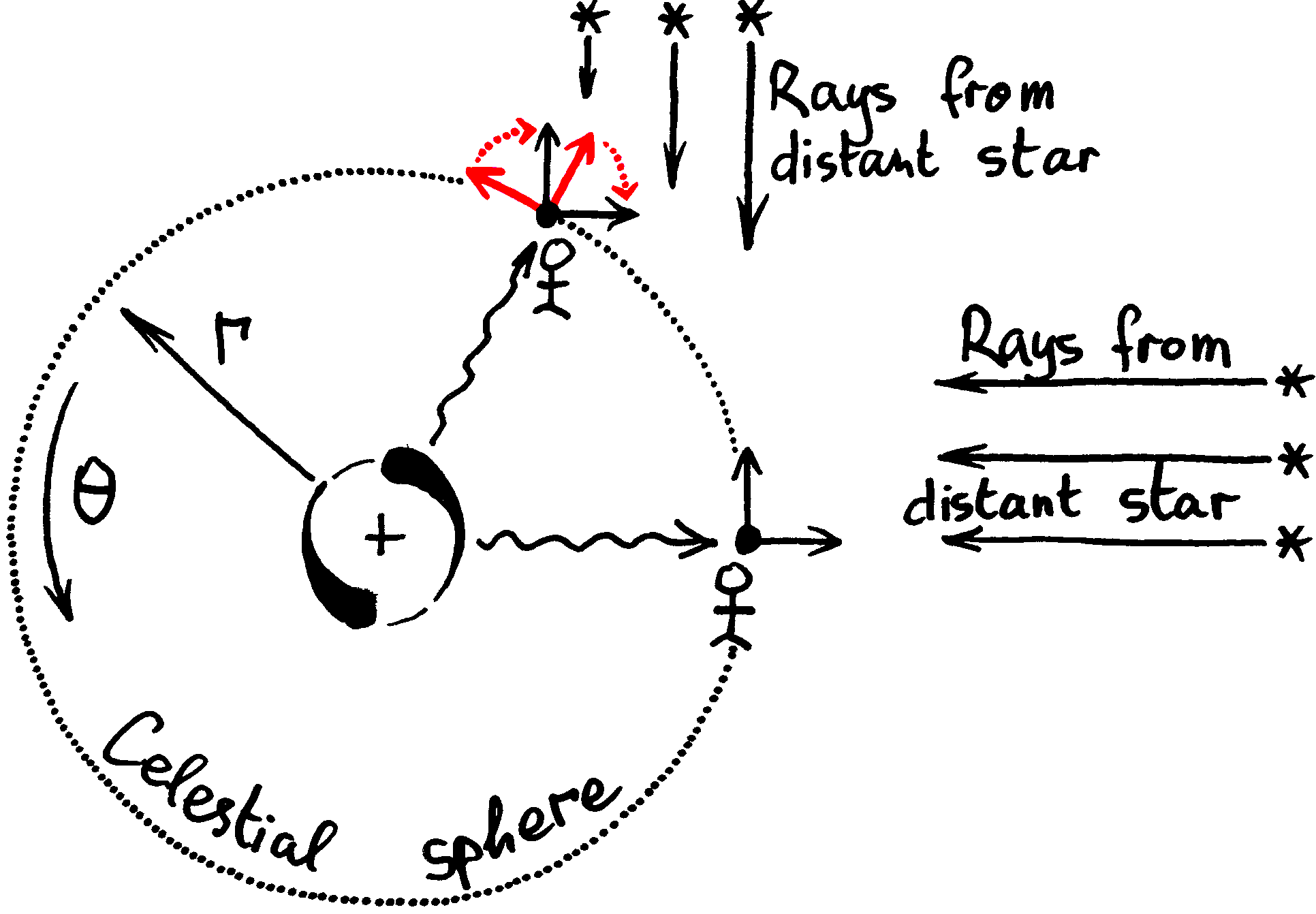}
\caption{An observer looks at a source at the origin. The observer's local frame (in black, on the far right) is initially built so that one of its axes coincides with the source's direction, and it also happens to point towards some distant star. As the observer moves, the source-oriented tetrad rotates so as to remain directed towards the source (red frame). Reorienting the tetrad so that its (black) axes point once again towards distant stars requires a (red) compensating rotation \eqref{e30}.}
\label{fiXX}
\end{SCfigure}

The set of vectors $\{\be{0},\be{r},\be{a}\}$ is thus a source-oriented Lorentz tetrad; it is good enough for many applications, including radiative effects at order $\cO(r^{-1})$, and one may compute the corresponding spin connection \eqref{e18} to deduce a precession rate \eqref{e19}. However, the source-oriented frame has a drawback for our purpose: it mixes the dynamical precession caused by gravitational waves with a kinematical precession due to the motion of the gyroscope on the asymptotic sphere. The source of this kinematic effect is that source-oriented frames have to rotate continuously, regardless of the presence of radiation, simply to align the radial basis vector towards the source (see fig.\ \ref{fiXX}). As we now show, the dynamical precession caused by gravitational waves can be singled out by transforming to a new frame `tied to distant stars', whose existence is guaranteed by the asymptotically flat nature of spacetime.

\subsection{Frame tied to distant stars}
\label{sestar}

The aforementioned drawback in the source-oriented frame can be explained easily: one expects free gyroscopes in Minkowski spacetime to experience no precession, but the spin connection \eqref{e18} of the origin-oriented frame in flat space ($F=1$, $\beta=0$, $U^a=0$, $\gamma_{ab}=q_{ab}$) reads
\beq
\label{e27}
\overline{\omega}_{\hr\ha}
=
-\overline{\zeta}_{\ha a}d\th^a,
\qquad
\overline{\omega}_{\ha\hb}
=
\barE{a}{b}D_a\overline{\zeta}_{\hb b}d\th^a,
\ee
where bars denote quantities evaluated in a Minkowskian background. This is manifestly nonzero: a free gyroscope with angular velocity $v^a$ precesses at a rate $-v^a\overline{\omega}_{a\hi\hj}(\th^a)$ with respect to the source-oriented frame, purely due to the fact that the tetrad needs to rotate continuously in order to keep pointing towards the origin. It would be much more convenient to devise a tetrad whose spin connection vanishes in flat space. In fact, in pure Minkowski space, the solution simply consists in using the Cartesian frame whose spin connection is indeed zero. However, the proper frame is not as obvious for generic asymptotically flat spacetimes, where one can at best define an \textit{asymptotically} inertial structure. One thus needs to build a frame tied to `distant stars' whose distance from the observer is effectively infinite, contrary to the source, which is at large but finite distance. 

\paragraph{Local rotation.} In practice, a frame $\{\bff{\mu}\}$ tied to distant stars can be obtained from the source-oriented tetrad $\{\be{\mu}\}$ by a celestially local rotation cancelling the spurious time-dependent rotation of the $\be{\mu}$'s. Thus we declare that
\beq
\label{e28}
\bff{0}\equiv\be{0}=\bu,
\qquad
\bff{i}\equiv R_{\hi}{}^{\hj}(\bth)\be{j},
\ee
where the rotation matrix $R_{\hi}{}^{\hj}(\bth)$ only depends on angular coordinates $\th^a$, is independent of the dynamical bulk metric, and may be chosen to reduce to the identity at some reference point $\bth_0$ to be thought of as the initial angular position of the observer. The change of tetrad \eqref{e28} entails a modification $\omega\to\omega'$ of the spin connection \eqref{e18}, expressed in terms of one-forms by the standard transformation law
\beq
\label{e29}
\omega^{\prime}_{\hi\hj}
=
R_{\hi}{}^{\hat k} R_{\hj}{}^{\hat\ell}\omega_{\hat k\hat\ell}
+R_{\hi}{}^{\hat k}dR_{\hj\hat k}.
\ee
Accordingly, the rotation that yields a frame whose spin connection $\omega'$ vanishes in Minkowski space is given by a path-ordered exponential
\beq
\label{e30}
R(\bth)
=
P\exp\int_{\bth_0}^{\bth}\overline{\omega},
\ee
where $\bar\omega$ is given by \eqref{e27} and should be seen as an $so(3)$-valued one-form. The curve connecting the reference point $\bth_0$ to $\bth$ is, in principle, arbitrary, and its choice affects the value of $R$ since the curvature of $S^2$ does not vanish. The choice is nevertheless ultimately irrelevant, as one merely has to pick \textit{some} path for every pair $(\bth_0,\bth)$ in a suitable open neighbourhood of the observer's initial position.\footnote{There exists no \textit{globally} well-defined rotation mapping the source-oriented tetrad on a frame tied to distant stars because the second homotopy group of $S^2$ is nontrivial. This is not an issue for our purposes, as we only need to cover some open subset of $S^2$ in which the observer's motion takes place.} We shall assume that one such choice has been made, in such a way that $R(\bth_0)=\mathbb{I}$ is the identity matrix.

The geometric justification of this construction goes as follows (see fig.\ \ref{fiXX}). Suppose the source-oriented frame, centred at $\bth_0$ at some arbitrary initial time $u_0$, points towards three stars at infinity, one of them aligned with the source's direction. Then let time flow, whereupon the source-oriented frame rotates in a time-dependent manner so as to keep pointing towards the source. As it does so, the observer's position on a celestial sphere also changes from $\bth_0$ to some $\bth$. Now let the observer perform a rotation \eqref{e30} to their frame at every step of their motion; doing so reorients the frame so that its vectors are aligned once more with the three stars chosen at time $u_0$, regardless of the (possibly dynamical) bulk metric or of other time-dependent quantities. Accordingly, we refer to the tetrad \eqref{e28} as being `tied to distant stars' at infinity.

It is worth noting that the implementation of such a tetrad in actual experiments is not far-fetched. For instance, the gyroscopes carried by Gravity Probe B \cite{Everitt:2011hp,Everitt:2015qri} were designed in just such a way that the dragging of inertial frames with respect to distant stars (\eg IM Pegasi) could be observed---albeit in the stationary metric of Earth. The use of similar methods in time-dependent, radiative metrics would undoubtedly present formidable technical challenges; we briefly return to such practical issues---and observational prospects---in the \hyperref[se5]{conclusion} of this work.

\paragraph{Spin connections.} The spin connection of a tetrad tied to distant stars is readily deduced from the transformation law \eqref{e29}. Since the rotation $R$ was chosen so as to annihilate the spin connection of Minkowski space ($R\overline{\omega}R^{-1}+RdR^{-1}=0$), one finds indeed
\beq
\label{e31}
\omega^{\prime}_{\hi\hj}
=
R_{\hi}{}^{\hm} R_{\hj}{}^{\hn}\big(\omega_{\hm\hn}-\overline{\omega}_{\hm\hn}\big),
\ee
where $\omega$ is the spin connection \eqref{e18} of the source-oriented frame built in section \ref{sesof}. The actual components of $\omega$ and $\omega'$ depend on one's choice of worldline---either a geodesic or a static line at fixed $(r,\bth)$, as mentioned at the end of section \ref{segyk}. It is therefore time to display these results: using the velocity \eqref{e14}--\eqref{gea} of geodesics at large $r$ and the ensuing source-oriented frame \eqref{e23}--\eqref{e24}, the spin connection \eqref{e18} follows after a long and painful, but otherwise straightforward calculation. The rotation \eqref{e31} then cancels the spurious change of orientation due to the leading angular velocity in \eqref{gea}, and one eventually finds\\ 
\begin{minipage}{.04\textwidth}
\rotatebox{90}{(Free fall)~~~}
\end{minipage}
\begin{minipage}{.95\textwidth}
\vspace{-.2em}
\begin{align}
\label{e33}
\omega^{\prime}_{\hr\ha}
&\sim
\barE{a}{a}\left(\frac{1}{4r^2} N_{ab}D_cC^{bc}du
+\frac{1}{2r^2}D^bC_{ab}dr
+\frac{1}{2}N_{ab}d\th^b\right),\\
\label{e34}
\omega^{\prime}_{\ha\hb}
&\sim
-\frac{1}{2r^2}\barE{a}{a}\barE{b}{b}\big(D_{[a}D^cC_{b]c}-\frac12 N_{c[a}C^c{}_{b]}\big)du+\cO(r^{-3})dr+\cO(r^{-1})d\th^a
\end{align}
\end{minipage}\\[.7em]
up to subleading corrections in $1/r$ that are implicitly neglected. (Our convention for antisymmetrization is $A_{[a}B_{b]}\equiv\frac12 (A_aB_b-A_bB_a)$.) This is expressed here in terms of on-shell metric data appearing in eq.\ \eqref{e13}, specifically in terms of the asymptotic shear $C_{ab}$ and the news tensor $N_{ab}=\pd_uC_{ab}$. Also note that we fix $C_{ab}(u=-\infty)=0$ as an initial condition: this slightly simplifies the spin connection by ensuring that the `$\Delta$' of the factors $\Delta C$ that used to affect the geodesic \eqref{geu}--\eqref{gea} may now be omitted. (Mass and angular momentum aspects do not contribute to the spin connection at this order in $1/r$; in particular, Lense-Thirring precession \cite[\S40.7]{Misner:1973prb} is hidden in subleading terms that are unimportant for us.)

The computation is identical for worldlines at constant $(r,\bth)$, save for the fact that proper velocity \eqref{e21} simplifies to $\bu=\gamma\pd_u$ (instead of the more involved expressions \eqref{e14}--\eqref{gea}). Using once more the tetrad \eqref{e23}--\eqref{e24} and the source-oriented spin connection \eqref{e18}, the rotated spin connection \eqref{e31} for a static worldline at constant $(r,\bth)$ reads\\
\begin{minipage}{.04\textwidth}
\rotatebox{90}{(Static)~~~}
\end{minipage}
\begin{minipage}{.95\textwidth}
\vspace{-.2em}
\begin{align}
\label{e35}
\omega^{\prime}_{\hr\ha}
&\sim
\barE{a}{a}\left(\frac{1}{2r} D^b N_{ab}du
+\cO(r^{-3})dr
+\frac{1}{2}N_{ab}d\th^b\right),\\
\label{e36}
\omega^{\prime}_{\ha\hb}
&\sim
-\frac{1}{2r^2}\barE{a}{a}\barE{b}{b}\big(D_{[a}D^cC_{b]c}-\frac12 N_{c[a}C^c{}_{b]}\big)du
+\cO(r^{-3})dr
+\cO(r^{-1})d\th^a,
\end{align}
\end{minipage}\\[.7em]
again up to neglected subleading corrections. Note the identical $\hat{a}\hat{b}$ components of the spin connection \eqref{e36} and its free-fall analogue \eqref{e34}. By contrast, the $\hat{r}\hat{a}$ components \eqref{e35} differ sharply from their geodesic cousin \eqref{e33}: the $u$ component of the former scales as $1/r$ at large $r$, while that of the latter goes as $1/r^2$. We return to this in section \ref{sestat}.

Eqs.\ \eqref{e33}--\eqref{e36} are crucial results for the remainder of this paper. They determine the precession rate \eqref{e19} according to $\Omega_{\hi\hj}=-u^{\mu}\omega^{\prime}_{\mu\hi\hj}$, where $\bu$ is either the geodesic velocity \eqref{e14}--\eqref{gea} or the static velocity $\bu=\gamma\pd_u$, respectively requiring the free-fall connection \eqref{e33}--\eqref{e34} or its static analogue \eqref{e35}--\eqref{e36}. The next section will indeed be devoted to a detailed study of this precession rate and its implications in terms of gravitational memory. (Note that one can, in fact, use a simplified expression for the precession rate: observers initially at rest in Bondi coordinates have an angular velocity $v^a=\cO(r^{-2})$, so the rotation matrix \eqref{e30} evaluated along the angular position of an observer satisfies $R=\mathbb{I}+\cO(r^{-2})$. It follows that $\Omega_{\hi\hj}\sim-u^{\mu}(\omega_{\mu\hi\hj}-\overline{\omega}_{\mu\hi\hj})$ at large $r$.)

\section{Gravitational memory from gyroscopic precession}
\label{se4}

The spin connections \eqref{e33}--\eqref{e36} determine the angular velocity \eqref{e19} of a spinning gyroscope, located far away from a gravitational source, with respect to a frame whose axes are tied to distant stars. We now investigate this precession in depth, mostly focussing on freely-falling observers. Then, the gyroscope's motion takes place in a plane orthogonal to the source's direction and turns out to be related to a suitable notion of dual mass aspect \cite{Freidel:2021qpz}. Bursts of gravitational radiation thus lead to net changes of orientation that may be interpreted as a memory effect. The latter turns out to involve a nonlocal superrotation charge and a flux term for angular momentum that stem from the balance equations \eqref{balance} for gravitational data, plus a duality generator. In the case of \textit{static}, accelerated observers, the leading precession stems from the $u$ component of \eqref{e35} and yields a memory effect that contains the same information as standard displacement memory \cite{Zeldovich:1974gvh,Braginsky:1985vlg,braginsky1987gravitational}; the notion of spin memory \cite{Pasterski:2015tva} is also recovered as a limit.

\subsection{Precession as a dual mass aspect}

Consider a freely-falling observer with velocity \eqref{e14}--\eqref{gea}, carrying a gyroscope whose precession rate \eqref{e19} involves the spin connection \eqref{e33}--\eqref{e34}. From this one readily finds our main result:
\begin{empheq}[box=\othermathbox]{equation}
\label{s11s}
\Omega_{\ha\hr}
=
\cO(r^{-3}),
\qquad
\Omega_{\ha\hb}
=
\frac{\eps_{\ha\hb}}{r^2}
\tl\cM
+\cO(r^{-3}),
\qquad
\tl\cM\equiv \frac14 D_{a}D_b\tl C^{ab}-\frac18 N_{ab}\tl C^{ab},
\end{empheq}
where $\eps_{\ha\hb}$ is the usual Levi-Civita symbol with two indices and the dual of any symmetric tensor $X_{ab}$ is defined as $\tl X_{ab}\equiv\eps_{c(a}X_{b)}{}^{c}=\eps_{ca}X_{b}{}^{c}-\frac{1}{2}\eps_{ab}X_c{}^c$ in terms of the Levi-Civita tensor density on $S^2$ \cite{Godazgar:2018dvh} (symmetrization is defined as $A_{(a}B_{b)}\equiv\tfrac{1}{2}(A_aB_b+A_bB_a)$ and may be dropped when $X$ is traceless). In particular,
\beq
\label{dush}
\tl C_{ab}
\equiv
\eps_{ca}C_{b}{}^{c}
\ee
is the \textit{dual shear tensor} that will play a key role below. 

Eqs.\ \eqref{s11s} state that, at leading order in the $1/r$ expansion, the gyroscope's axis only rotates in the plane tangent to the celestial sphere at the observer's location. The precession frequency is set by the quantity $\tl\cM(u,\bth)$, dubbed `dual covariant mass aspect' \cite{Freidel:2021qpz} in analogy with the usual covariant mass aspect $m+\frac18 N_{ab} C^{ab}$ \cite{Compere:2018ylh,Donnay:2021wrk}.\footnote{This justifies the notation $\tl\cM$, as $\cM$ normally indicates the usual (non-dual) covariant mass aspect. We stress that $\tl\cM$ has nothing to do with mass as such: it is related, instead, to multipole moments of the gravitational angular momentum (see eq.\ \eqref{e42} where $C^-$ is written solely in terms of spin multipole moments, or eq.\ (4.8b) of \cite{Compere:2019gft}). Mass and spin are indeed mutually dual in general relativity.} Indeed, the latter reduces to $\tl\cM$ upon using the first balance equation in \eqref{balance} to rewrite $m$  in terms of the shear, then dualizing $C_{ab}$ into $\tl C_{ab}$. One may think of the linear term $D_aD_b\tl C^{ab}$ as the vorticity of the angular motion of the geodesic flow stemming from eq.\ \eqref{gea}; it coincides with the vorticity of the congruence of static worldlines at large $r$.\footnote{To be precise, $D_aD_b\tl C^{ab}$ is the \textit{leading term} of vorticity near null infinity. That both static worldlines and geodesics have the same leading-order vorticity stems from the asymptotic geodesic velocity $\bu\sim\der_u$.} As for the nonlinear term $N_{ab}\tl C^{ab}$, it turns out to be necessary for $\tl\cM$ to transform covariantly (without inhomogeneous terms) under Weyl-BMS transformations \cite{Freidel:2021qpz}. We will also see below that $N_{ab}\tl C^{ab}$ is essentially the generator of local duality on the gravitational phase space. Gyroscopic precession near null infinity thus provides an observational protocol for the dual covariant mass aspect, as the latter essentially coincides with the rate \eqref{s11s}. We now massage the expression of $\tl\cM$ so as to set the stage for the computation of memory in section \ref{segymem}.

\paragraph{Parity decomposition.} It is convenient, for later use, to Hodge-decompose tensors on the sphere in terms of scalar quantities with definite parity. This will indeed allow us to relate the linear term of \eqref{s11s} to superrotation charges in section \ref{segymem}. Accordingly, in what follows we write angular momentum and shear as
\beq
\label{e41}
L_a
\equiv
D_aL^++\eps_{ab}D^bL^-,
\qquad
C_{ab}
\equiv
D_{\langle a}D_{b\rangle} C^{+}+\epsilon_{c(a} D_{b)} D^{c} C^{-},
\ee
where superscripts $\pm$ refer to parity eigenvalues of the respective functions, while angular brackets denote the symmetric, trace-free projection $D_{\langle a} D_{b\rangle}\equiv\tfrac{1}{2}(D_aD_b+D_bD_a)-\tfrac{1}{2}q_{ab} D^2$. (Without loss of generality, we shall assume that $L^\pm$ have no zero mode, while $C^\pm$ have no $\ell=0,1$ harmonics; this will be useful below when `inverting' the decomposition \eqref{e41}.) Using this, one finds that the linear term in the precession rate \eqref{s11s} has odd parity:
\beq
\label{e42}
\tl\cM
=
\frac18D^2(D^2+2)C^--\frac18N_{ab}\tl C^{ab}.
\ee
The balance equation \eqref{balance} for angular momentum now implies the relation $\tfrac{1}{8}D^2(D^2+2)C^-=\dot L^-+\cJ^-$, where the pseudoscalar flux $\cJ^-$ is defined analogously to the decomposition \eqref{e41} for angular momentum and similarly has no zero mode by definition. It follows that the dual mass aspect \eqref{e42} can finally be recast as
\beq
\label{e43}
\tl\cM
=
\dot L^-
+\cJ^-
-\frac18N_{ab}\tl C^{ab}.
\ee
A corollary of this rewriting is the absence of precession in nonradiative spacetimes, where the angular momentum aspect is constant and fluxes vanish since they are proportional to the news. We now exploit the decomposition \eqref{e43} to relate gyroscopic memory to certain symmetries of the gravitational phase space.

\subsection{Gyroscopic memory and gravitational symmetries}
\label{segymem}

The gyroscope carried by a freely-falling observer precesses according to eq.\ \eqref{e19}, with a rate \eqref{s11s} that vanishes in the absence of radiation thanks to the rewriting \eqref{e43} of the dual mass aspect. It follows that any finite burst of gravitational radiation leaves a permanent imprint---a memory---of its passage on the gyroscope's orientation. To compute this effect, our starting point is the precession equation \eqref{e19}, whose solution can formally be written as a time-ordered exponential of $\Omega$ acting on the initial spin vector. Since the angular velocity \eqref{s11s} is only accurate up to order $\cO(r^{-2})$, the expansion of the exponential is only reliable up to first order in $\Omega$. The net change of orientation due to gravitational waves is therefore
\beq
\label{e45}
\Delta S^{\hr}
=
\cO(r^{-3}),
\qquad
\Delta S^{\ha}
=
\Phi\,\eps^{\ha\hb}S^{\text{initial}}_{\hb}
+\cO(r^{-3}),
\ee
where the net rotation angle $\Phi=\frac{\overline{\Phi}}{r^2}$ decays as the square of the inverse distance to the source and involves the crucial radius-independent factor
\beq
\label{gymemo}
\overline\Phi
\equiv
\frac14 D_{a}D_b\int du\,\tl C^{ab}-\frac18 \int du\, N_{ab}\tl C^{ab}\,.
\ee
(Henceforth, we write $\Delta X\equiv X(u=+\infty)-X(u=-\infty)$ and all time integrals run from $u=-\infty$ to $u=+\infty$.) The gyroscopic memory effect thus consists of a `soft' part linear in the shear and a `hard' quadratic part, respectively given by the first and second terms of \eqref{gymemo}. We now discuss these effects, justifying the nomenclature along the way, and end by comparing the precession rate \eqref{s11s} to asymptotic rotations of the dyad \eqref{dyad-bulk} \cite{Godazgar:2022foc}.

\paragraph{Linear gyroscopic memory.} The linear part of gyroscopic memory \eqref{gymemo} coincides with the spin memory effect \cite{Pasterski:2015tva}, which can be formulated in terms of superrotation charges and fluxes \cite{Nichols:2017rqr}. Indeed, the first two terms on the right-hand side of \eqref{e43} allow us to write
\begin{align}
\label{phisoft}
\overline{\Phi}_{\text{soft}}
&\equiv
\frac{1}{4}
D_aD_b\int du\,\tl C^{ab}=\Delta L^-+\int du\,\cJ^-
\end{align}
where $\Delta L^-\equiv\int du\,\dot L^-$ is the net change of the parity-odd part of the angular momentum aspect of the source, and $\int du\,\cJ^-$ is the corresponding flux carried by hard gravitational radiation. To relate this to charges of asymptotic symmetries, note from the Hodge decomposition \eqref{e41} that $L^-=D^{-2}(\eps^{ab}D_bL_a)$ where $D^{-2}$ is the inverse of the Laplacian on the sphere, given by the Green's function 
\begin{align}
\label{Green}
D^2G(\bth,\bth')=\frac{1}{\sqrt{q(\bth)}} \delta^{2}(\bth-\bth')-\frac{1}{4 \pi},
\qquad
G(\bth,\bth')=\frac{1}{4 \pi} \log \sin ^{2} \frac{|\bth-\bth'|}{2}
\end{align}
with $|\bth-\bth'|$ the geodesic distance between the points $\bth$ and $\bth'$ on the round sphere. (The Green's function can be derived by exploiting rotational symmetry to turn the Poisson equation in \eqref{Green} into an ordinary differential equation: see \eg \cite[\S4.2]{aubin1998some} for details.) Accordingly, the soft memory contribution \eqref{phisoft} can be written as
\begin{empheq}[box=\othermathbox]{equation}
\label{e47}
\overline{\Phi}_{\text{soft}}
=
8\pi\big(\Delta Q_Y+\cF_Y\big)
\end{empheq}
where $Y^a(\bth')\equiv\eps^{ab}D'_bG(\bth,\bth')$ is a divergence-free vector field on $S^2$, depending parametrically on $\bth$, whose superrotation charge $Q_Y$ and flux $\cF_Y$ are given by\footnote{There is a subtlety here: one can redefine the charge and flux \eqref{e48} by mapping $L_a\to\widehat{L}_a=L_a+\alpha_a$ and $\cJ_a\to \widehat{\cJ}_a=\cJ_a-\dot{\alpha}_a$ for any one-form $\alpha_a$, and using hatted quantities in \eqref{e48}. For the charge to match its Wald-Zoupas version \cite{Wald:1999wa}, one has to take $\alpha_a=-u D_{a} m-\frac{1}{16} D_{a}\left(C_{bc} C^{bc}\right)-\frac{1}{4} C_{ab} D_{c} C^{bc}$ \cite{Flanagan:2015pxa}. Fortunately, the sum $\Delta Q_{Y}+\cF_Y$ in \eqref{e47} is unaffected by this ambiguity.}
\beq
\label{e48}
Q_{Y}
\equiv
\frac{1}{8\pi}
\oint_{S^2}
\sqrt{q}\,d^2\bth'
\,Y^a(\bth')\,L_a(\bth'),
\qquad
\cF_Y
\equiv
\frac{1}{8\pi}
\int du
\oint_{S^2}
\sqrt{q}\,d^2\bth'
\,Y^a(\bth')\,\cJ_a(\bth').
\ee
These expressions now involve the usual angular momentum aspect $L_a$ and its flux $\cJ_a$ defined in \eqref{fluxes}, without parity-odd projection; the projection is actually automatic (\ie only pseudoscalars contribute) thanks to the parity-odd form of the vector field $Y^a$. 

\paragraph{Nonlinear gyroscopic memory.} Let us now discuss the nonlinear term of the gyroscopic memory effect \eqref{gymemo}:
\begin{empheq}[box=\othermathbox]{equation}
\label{exof}
\overline{\Phi}_{\text{hard}}
\equiv
-\frac{1}{8}\int du\,N_{ab}\tl C^{ab}.
\end{empheq}
We will show that this quantity has several neat interpretations: it is the generator of local gravitoelectric-gravitomagnetic duality on celestial spheres, and it may also be seen as the local helicity of passing waves. Indeed, recall first that the linearized Einstein equations are invariant under infinitesimal duality transformations $\delta R_{\mu\nu\alpha\beta}=\varepsilon \star R_{\mu\nu\alpha\beta}$, where $\star$ denotes the Hodge dual and $\varepsilon$ is a small parameter \cite{Henneaux:2004jw}. The corresponding transformation of Bondi shear at null infinity reads $\delta C_{ab}=\varepsilon\tl C_{ab}$, where $\tl C_{ab}$ is the dual shear \eqref{dush}. For our purposes, it will be of interest to consider \textit{local} duality transformations on celestial spheres, namely
\begin{align}
\label{duality}
\delta C_{ab}=\varepsilon(\bth)\tl C_{ab}\,,\qquad\delta N_{ab}=\varepsilon(\bth)\tl N_{ab}
\end{align}
where $\varepsilon(\bth)$ is any smooth function on $S^2$. These transformations turn out to be canonical: they preserve the symplectic form on radiative phase space \cite{Ashtekar:1981bq},
\beq
\label{symplectic structure}
\Omega
=
\frac{1}{32\pi}
\int du
\oint_{S^2}
\sqrt{q}\,d^2\bth
\,\delta N^{ab}\wedge\delta C_{ab},
\ee
in the sense that the functional Lie derivative $\cL_{\delta_{\varepsilon}}\Omega=\delta[\Omega(\de_{\veps},\cdot)]=0$ vanishes.\footnote{The normalization of the symplectic form \eqref{symplectic structure} is merely chosen for convenience here, and differs from the standard normalization that involves Newton's constant owing to the Einstein-Hilbert action.} One can thus build the corresponding canonical generator $H$ such that $\delta H=\Omega(\delta_{\varepsilon},\cdot)$, which yields $H=\tfrac{1}{4\pi}\oint\sqrt{q}\,d^2\bth\,\varepsilon\,\overline{\Phi}_{\text{hard}}$ in terms of the quadratic memory \eqref{exof}. As announced, this confirms that nonlinear gyroscopic memory involves the generator of electric-magnetic duality.

We now turn to the link between the nonlinear memory \eqref{exof} and the helicity of gravitational radiation. This requires the mode expansion of the Bondi shear, obtained from the null-infinity limit of the solution of the linearized Einstein equations. Namely, following the conventions of \cite[eqs.\ (5.4)--(5.10)]{He:2014laa} and using natural units in which $G=1$, one has
\beq
\label{shearr}
C_{ab}(u,\bth)
=
\frac{\sqrt{32\pi}}{(2\pi)^3}
\lim_{r\to\infty}
\frac{1}{r}\frac{\der x^{\mu}}{\der\th^a}\frac{\der x^{\nu}}{\der\th^b}
\sum_{\alpha=1,2}
\int\frac{d^3{\bq}}{2|{\bq}|}
\Big[
\epsilon^{\alpha}_{\mu\nu}({\bq})^*\,
\hat a_{\alpha}({\bq})\,
e^{-i|{\bq}|u-i|{\bq}|r(1-\bn\cdot\bn')}+\text{h.c.}
\Big].
\ee
Here $\bn'\equiv\bq/|\bq|$ and $\bn$ is the unit normal vector at the observation point $\bth$,  while $\hat a_{\alpha}({\bq})^{(\dagger)}$ is the annihilation (creation) operator for a graviton with momentum ${\bq}$ and polarization tensor $\epsilon^{\alpha}_{\mu\nu}({\bq})$, with $\alpha$ labelling its polarization. The integration measure $d^3{\bq}/2|{\bq}|$ on the mass shell is Lorentz-invariant, so the commutators of Fock space operators read
\beq
\label{commutator}
\big[\hat a_{\alpha}({\bq}),\hat a_{\beta}({\bk})^{\dagger}\big]
=
\delta_{\alpha\beta}\,2|{\bq}|(2\pi)^3\,\delta^{3}({\bq}-\bk).
\ee
The large-$r$ limit in \eqref{shearr} crucially localizes the momentum integral to the saddle point where ${\bq}$ points along the observation direction $\bth$ \cite{He:2014laa}.\footnote{This is actually a group-theoretic statement that relates massless representations of Poincar\'e to those of BMS: see \eg \cite[\S2.4]{Bekaert:2022ipg} for details and references.} As a result, one can rewrite the shear as
\beq
\label{shearb}
C_{ab}(u,\bth)
=
\frac{i\pi\sqrt{32\pi}}{(2\pi)^3}
\sum_{\alpha=1,2}
\int_0^{\infty}d\omega
\Big(f^{\alpha}_{ab}(\bth)^*\,
\hat a_{\alpha}(\omega,\bth)\,
e^{-i\omega u}
-
\text{h.c.}\Big)
\ee
where $f_{ab}^{\alpha}(\bth)\equiv\frac{1}{r^2}\der_ax^{\mu}\der_bx^{\nu}\eps_{\mu\nu}^{\alpha}(\bth)$ maps the polarization tensor to a symmetric-trace free tensor on the celestial sphere. It is then a simple matter to write the integral \eqref{exof} as a Fock space operator; after some algebra involving polarization tensors, one finds
\begin{align}
\overline{\Phi}_{\text{hard}}
&=\frac{1}{2\pi^2}i\eps^{\alpha\beta}\int_0^\infty d\omega \,\omega \, a_\alpha a^\dagger_\beta
=
\frac{1}{2\pi^2}\int_0^\infty d\omega \,\omega \, (a^\dagger_+a_+-a^\dagger_-a_-),
\end{align}
where the second equality was obtained in a complex dyad (helicity basis) in which $\eps^{\alpha\beta}=\text{diag}(-i,i)$. This is the result announced above: the nonlinear memory measures the net helicity of the wave burst that has crossed the gyroscope's worldline, \ie the difference between the numbers of left- and right-handed gravitons that have crossed future null infinity at $\bth$.

\paragraph{Precession and residual frame symmetry.} Recall from section \ref{sesof} that the tetrad with angular vectors \eqref{dyad-bulk} is only fixed up to local U(1) rotations of the dyad $\barE{a}{}(\bth)$ on the celestial sphere. As it turns out, such transformations are (subleading) asymptotic symmetries of first-order general relativity \cite{Godazgar:2022foc} provided the Einstein-Hilbert action is supplemented by a (boundary) Gauss-Bonnet term. It is thus tempting to think that gyroscopic memory is an effect of vacuum transitions under such rotations, similarly to the link between displacement memory and supertranslations \cite{Strominger:2014pwa}. After all, the two effects involve analogous equations: displacement memory is a net change $\Delta X_\ha=\frac{1}{r}X_0{}^\hb\,\Delta C_{\ha\hb}$ in the separation $X^{\ha}\E{a}{}$ of nearby geodesics (with initial separation $X_0{}^\hb$), and gyroscopic memory \eqref{e45} reads $\Delta S_\ha=\frac{1}{r^2}S_{0}{}^\hb\,\eps_{\ha\hb}\,\overline{\Phi}$ in terms of the quantity \eqref{gymemo} discussed at length above. But several key differences lurk behind this superficial similarity. First, the change of shear that enters displacement memory can be seen as a genuine supertranslation, while the rotation \eqref{e45} of the gyroscope's spin is subleading compared to the asymptotic symmetry transformations in \cite{Godazgar:2022foc}. Second, the precession rate \eqref{s11s} is essentially the boundary Noether current of U(1) asymptotic symmetries (just like the Bondi mass aspect is the current of supertranslations), so the rotation \eqref{e45} cannot be interpreted as a net change in the value of U(1) surface charges. There are thus striking links between gyroscopic memory and the U(1) symmetries of \cite{Godazgar:2022foc}, but one should be careful when comparing them to the simpler relation between displacement memory and supertranslations.

\subsection{Comments on related works}
\label{sestat}

The study of gravitational-wave effects on spinning bodies is not new, going back at least to the early works \cite{Herrera:2000uh,Bini:2000xj,ValienteKroon:2001pc}. Following the discovery of a relation between displacement memory and BMS supertranslations \cite{Strominger:2014pwa}, memory effects involving angular momentum were similarly considered in \cite{Pasterski:2015tva} in terms of Sagnac interferometers, and in \cite{PhysRevD.84.124014,Flanagan:2014kfa,Flanagan:2018yzh} in terms of the relative precession of nearby gyroscopes. Here we shall briefly compare our construction above to two such results in the literature: the papers \cite{Herrera:2000uh,Herrera:2013poa} on vorticity in radiative spacetimes, and the spin memory effect of \cite{Pasterski:2015tva}.

A first general remark is that these references have one common aspect: they use \textit{static} observers at large $r$, as opposed to freely-falling ones. While this simplifies computations since the corresponding velocity is just $\bu=\gamma\der_u$, it complicates their comparison to experiments where a detector is typically expected to fall freely as radiation crosses its path---think \eg of the freely-falling Gravity Probe B experiment \cite{Everitt:2011hp,Everitt:2015qri}. Forcing the detector to remain static while being subjected to a dynamical metric requires delicately fine-tuned, time-dependent accelerations whose practical feasibility is questionable---with the tiniest of errors likely to affect the measurement of an already tiny observable.

Having said this, it is still true that the static-observer spin connection \eqref{e35}--\eqref{e36} contains valuable insights on gravitational radiation. Consider first the leading $\hr\ha$ part \eqref{e35}, whose time component decays like $1/r$ as opposed to $1/r^2$ in \eqref{e36}. The corresponding precession rate is
\beq
\label{e49}
\Omega_{\ha\hr}
\sim
\frac{1}{2r}\barE{a}{a} D^b N_{ab},
\ee
so that the main rotation of gyroscopes carried by static observers occurs along an axis tangent to the celestial sphere. As it turns out, this effect was first found and discussed in \cite{Herrera:2000uh,Bini:2000xj,ValienteKroon:2001pc} thanks to the vorticity of static worldlines in Bondi coordinates. Its radial order $\cO(r^{-1})$ could have been expected, since worldlines at constant $(r,\bth)$ have a nonzero acceleration $u^\nu\cd_\nu u^\mu\sim(\frac{\dot{m}}{r},-\frac{\dot{m}}{r},\frac{1}{2r^2}D_b N^{ab})$ at large $r$. From the perspective of gravitational memory, the net rotation due to \eqref{e49} is proportional to the parity-even variation of shear $\Delta C^+$ (recall the Hodge decomposition \eqref{e41}), so it contains the same information as standard displacement memory, determined by the balance equation for supertranslations \cite{Zeldovich:1974gvh,Braginsky:1985vlg,braginsky1987gravitational}. Amusingly, an analogous leading-order precession in the $\hr\ha$ plane occurs for magnetic dipoles near null infinity \cite{Oblak:2023axy}.

Let us now turn to the subleading time component of the static spin connection, namely the $\ha\hb$ part \eqref{e36}. As we have seen, this is the same contribution that gives rise to the leading precession \eqref{s11s} of freely-falling gyroscopes. But it can also be used to compute the effect of  gravitational waves on a Sagnac interferometer facing their source, which connects our work to the spin memory effect \cite{Pasterski:2015tva}. To see this, recall the Sagnac effect: when a light beam of wavelength $\lambda$ is split in two separate beams moving in opposite directions along some planar closed path $\cC$ (say an optical fibre), the apparatus's constant angular velocity $\Omega$ causes a relative phase shift \cite{wang2004generalized}
\beq
\label{e51}
\delta \psi
=
\frac{8\pi}{\lambda c}\int_{\text{Int}(\cC)}\Omega_{ij}\,dx^i\wedge dx^j
\ee
after one orbit, where $c$ is the speed of light that we momentarily reinstate. The phase accumulated over $n$ such periods is just $n\,\delta\psi$. Now suppose $\Omega$ depends on time and vanishes at early and late times, but varies very slowly so that $\Omega_{ij}(u)$ is nearly constant over the course of one period $\delta t=\ell/c$, where $\ell$ is the length of the loop $\cC$. Then the total Sagnac phase accumulated during the experiment is
\begin{align}
\label{depp}
    \Delta \psi
    =
    \int_{-\infty}^{+\infty}du\frac{\de\psi(u)}{\de t}
=
\frac{8\pi}{\lambda\ell}\int_{\text{Int}(\cC)}\int du\,\Omega_{ij}(u)\,dx^i\wedge dx^j.
\end{align}
This exactly applies to the gyroscopic setup of the present paper. Indeed, consider a Sagnac interferometer located far away from a gravitational source, and assume that the closed path $\cC$ lies in a plane transverse to the direction of wave propagation (\ie tangent to the sphere at some large radius). The precession rate \eqref{s11s} plugged into \eqref{depp} then reveals a net phase shift proportional to the gyroscopic memory \eqref{gymemo},
\begin{align}
\label{depth}
    \Delta \psi=\frac{4\pi R}{\lambda }\,\Phi,
\end{align}
where $R\equiv 2\,\text{area}(\cC)/\ell$ is the effective radius of the interferometer. Since gyroscopic memory contains a term \eqref{phisoft} linear in the Bondi shear, the phase shift \eqref{depth} is very similar to the spin memory effect \cite{Pasterski:2015tva} but crucially differs from it by the nonlinear term \eqref{exof}, \ie the net helicity of radiation. A more detailed comparison between our result and spin memory is therefore relegated to future investigations.

\section{Conclusion}
\label{se5}

This work was devoted to a seemingly elementary exercise in general relativity, with striking results. Simply put, we addressed the analogue of Lense-Thirring precession for freely-falling gyroscopes in radiative spacetime manifolds. The key subtlety was to choose a tetrad tied to distant stars whose existence is guaranteed by asymptotic flatness; the orientation of the gyroscope was then measured relative to that tetrad. As we have seen, this yields a precession rate \eqref{s11s} proportional to the covariant dual mass aspect $\tl\cM$ of Bondi metrics \cite{Freidel:2021qpz}; the problem is thus deeply related to symmetries of the gravitational phase space at null infinity. The same is true of the net memory effect \eqref{e47}, whose linear term reproduces the known expression of spin memory \cite{Pasterski:2015tva} while also containing a surprising new, nonlinear contribution, due to the generator of celestially local electric-magnetic duality.

These conclusions illustrate the predictive power of large-distance asymptotic symmetries in gauge theories and gravity, even beyond the strict limit where observers sit `at the boundary of spacetime'. Furthermore, a number of related questions deserve to be investigated. A first fundamental puzzle is the following: one could have guessed that the orientation of a gyroscope would be related to gravitomagnetic effects, hence to `dual charges', but the justification for the appearance of the specific covariant dual mass in \eqref{s11s} is mysterious. It would be pleasing to find a more intuitive argument that `explains' the appearance of this quantity in a precession equation. Relatedly, it may be of interest to compute the analogous spin precession in backgrounds that possess an explicit gravitomagnetic monopole---typically the Taub-NUT metric \cite{Taub:1950ez,Newman:1963yy}---where the celestial average of the precession rate is expected to be quantized. In fact, in that last case, the assumption that $L^-$ has no zero-mode in the decomposition \eqref{e41} fails to hold and the rewriting \eqref{e43} of the dual mass aspect generally involves a nonzero derivative $\dot L^-$ even in the absence of radiation. It is unclear if a well-defined notion of memory even exists in such a setup.

From a more practical perspective, a key issue is to understand the order of magnitude of the memory effect described here and the feasibility of its actual observation. The first thing to say is that the effect, unsurprisingly, is extremely weak: as explained at the end of the short companion paper \cite{Seraj:2022qyt}, elementary dimensional analysis predicts a total rotation angle
\beq
\label{omag}
\Phi
\sim
\frac{G^2}{c^4}\frac{M^2}{r^2}
\simeq
2\times10^{-39}\,\Big(\frac{M/M_{\odot}}{r/1\,\text{Mpc}}\Big)^2
\ee
where $M$ is some mass scale determined by the source of radiation, $M_{\odot}$ is the solar mass and $r$ is the distance between source and observer. This is indeed tiny for relatively `light' sources of radiation, but it grows substantially for supermassive black hole mergers where values such as $\Phi\simeq10^{-26}\,\text{rad}$ are conceivable. Unfortunately, even such an enhanced signal is too weak for the current precision of gyroscopes; a good point of comparison is given by the Gravity Probe B experiment \cite{Everitt:2011hp,Everitt:2015qri}, whose measurement of frame dragging involved precession rates of about $2\times10^{-7}\,\text{rad}/\text{year}$, corresponding to a total rotation angle of roughly $10^{-6}\,\text{rad}$. Improving these figures by the many orders of magnitude needed for gyroscopic memory seems difficult, to say the least.

But this is not the end of the story. After all, standard displacement memory suffers from the similar issue of its tininess, and it is still likely to be observed in the coming decade thanks to the combination of the many gravitational wave events currently being observed \cite{Lasky:2016knh,Boersma:2020gxx,Grant:2022bla}. An analogous method could be applied to gyroscopic memory by superimposing data on gyroscope orientations from a large number of detections---although one would then have to forego detailed quantitative checks in favour of merely qualitative assessments.\footnote{A weakness of this method is that it requires an immense amount of observed events: the signal-to-noise ratio grows as the square root of the number of events, so enhancing the signal \eqref{omag} by \eg a factor $10^{10}$ requires an unrealistically large number of sources (about $10^{20}$).} Alternatively, one may turn to the sky and use distant pulsars instead of gyroscopes near Earth: if such a pulsar happens to be located close enough to a source of gravitational waves, its rotation \eqref{omag} could be sizeable enough to be visible from Earth---at least provided the time scale of radiation is much longer than the pulsar's period \cite{Lorimer:2008se}. Observing effects of this kind would provide shining examples of general-relativistic dynamics at work.

\section*{Acknowledgements}

We are grateful to Glenn Barnich, Jorrit Bosma, Marc Geiller, Mahdi Godazgar, Gregory Kozyreff and Marios Petropoulos for illuminating interactions on (dual) asymptotic symmetries, precession, and memory effects. In addition, we thank Miguel Paulos for suggesting to use pulsars as gyroscopes, as well as Kaye Jiale Li and Joana Teixeira for illuminating interactions on pulsars. A.S.\ also wishes to thank Gilles Esposito-Far\`ese, Roberto Oliveri and Simone Speziale for crucial discussions on gyroscopes and memory. Finally, we thank the anonymous referee for a manifestly careful reading of our manuscript and questions that contributed to an improved updated paper.

The work of A.S.\ is funded by the European Union’s Horizon 2020 research and innovation programme under the Marie Sk{\l}odowska-Curie grant agreement No.\ 801505. The work of B.O.\ is supported by the ANR grant \textit{TopO} No.\ ANR-17-CE30-0013-01, and by the European Union's Horizon 2020 research and innovation programme under the Marie Sk{\l}odowska-Curie grant agreement No.\ 846244.

\addcontentsline{toc}{section}{References}

\providecommand{\href}[2]{#2}
\end{document}